\documentstyle{article} 
\title{On partially formal supermanifolds.}
\author{Anatoly Konechny and Albert Schwarz \\
Department of Mathematics, University of California, \\
Davis, CA 95616 \\
konechny@math.ucdavis.edu, \\
schwarz@math.ucdavis.edu }

\begin{document}
\maketitle

\smallskip

\begin{abstract}
We define a finite-dimensional partially formal supermanifold as a manifold having $q$ odd coordinates and 
$k + l$ even coordinates with  $l$ of them  taking only nilpotent values. 
We show that this notion can be used to formulate superconformal field theories with different numbers 
of supersymmetries in holomorphic and antiholomorphic sectors. 
\end{abstract}

The present preprint is a new version of the paper\cite{OO} published as a preprint in 1997. This paper  and the companion paper \cite{0} stem from an attempt to understand the precise mathematical meaning of some 
constructions used by physicists. 
We have in mind in particular the consideration of independent 
spin structures in holomorphic and antiholomorphic sectors of N=1 superconformal field theory  and 
the notion of chiral (heterotic) supermanifold  (for example see \cite{MNP}). 
We introduce the notion of 
a partially formal supermanifold (a manifold that has in addition to standard even and odd coordinates also even nilpotent coordinates) and show that in many cases rigorous definitions can be based on 
this notion. Mathematical details are relegated to appendices. 
The appendices also contain the details of the formulation of supergeometry in terms of functors.
This language is most suitable for our purposes and is convenient 
for many other supergeometry questions as well. 

We came back to our old paper because the language of functors and of partially formal supermanifolds could be useful in the new approach to string theory suggested in \cite{GAS}. 

In the old version of the paper, we used the term $(k\oplus l, q)$-dimensional manifold for a manifold with $q$ odd and $k+l$ even coordinates where $l$ coordinates are nilpotent.  In the present version, we consider also an infinite-dimensional situation where this terminology is not appropriate  (the infinite-dimensional case was considered also in \cite {0}). 

The definitions that can be used in infinite-dimensional cases are given in Appendix A.

Let us start with the definition of superspace in terms of the space of $\Lambda$-points.

Let $\Lambda=\Lambda_{0} + \Lambda_{1}$ be a Grassmann algebra with an even subspace $\Lambda_{0}$ and 
an odd subspace $\Lambda_{1}$. 
The space ${\bf R}_{\Lambda}^{p|q}$ can be defined as a space consisting of rows 
$(x^{1}, \dots , x^{p}, \xi^{1}, \dots , \xi^{q})$ where $x^{1}, \dots , x^{p}\in \Lambda_{0}$ are even elements 
of Grassmann algebra $\Lambda$ and  $\xi^{1}, \dots , \xi^{q}\in \Lambda_{1}$ are odd elements from $\Lambda$. 
Physicists usually say that one can take as $\Lambda$ any Grassmann algebra provided it is large enough. 
From the viewpoint of a mathematician, it is better to consider a family of sets ${\bf R}_{\Lambda}^{p|q}$ 
corresponding to all Grassmann algebras $\Lambda$. It is easy to see that a parity-preserving homomorphism 
$\alpha: \Lambda \rightarrow \Lambda'$ generates naturally a map 
$\tilde \alpha: {\bf R}_{\Lambda}^{p|q} \rightarrow {\bf R}_{\Lambda'}^{p|q}$ and that 
$\widetilde{\beta \alpha} =  \tilde \beta \tilde \alpha$ for any parity preserving homomorphisms 
$\alpha: \Lambda \rightarrow \Lambda' , \enspace  \beta:  \Lambda' \rightarrow \Lambda''$. 
In the language of mathematics, this means that the correspondence 
$\Lambda \mapsto  {\bf R}_{\Lambda}^{p|q}$ determines
  a functor acting from the category of Grassmann algebras into the category of sets (or of vector 
spaces).

An (even) superfield on ${\bf R}^{p|q}$ can be defined as an expression of the form 
\begin{equation}
\sum_{k=2l} \sum_{1\le i_{1} < \dots < i_{k}}f_{i_{1},\dots ,i_{k}}(x^{1},\dots , x^{p})\xi^{i_{1}}\dots \xi^{i_{k}}
\label{ber}
\end{equation}
Here $f_{i_{1},\dots ,i_{k}}$ are smooth functions on ${\bf R}^{p}$. It is important to notice that such an expression 
determines a map 
$F_{\Lambda}:{\bf R}_{\Lambda}^{p|q} \rightarrow \Lambda_{0}={\bf R}_{\Lambda}^{1|0}$. 
This follows from 
the fact that we can substitute an even element of a Grassmann algebra into any smooth function of real variable. 
To verify this statement we notice that every even element $x$ of Grassmann algebra  can be represented 
in the form $x=m+n$ where $m\in {{\bf R}}$ and $n$ is nilpotent. We define $f(x)$ using the Taylor expansion 
$f(x)=\sum_{k=0}^{\infty}\frac{f^{(k)}(m)}{k!} n^{k}$ (the series terminates because $n$ is nilpotent). 
If $\alpha:\Lambda \rightarrow \Lambda'$ is a parity preserving homomorphism then 
$\tilde \alpha \circ F_{\Lambda} =  F_{\Lambda}\circ \tilde \alpha$.
 This means in  mathematical terminology  that a superfield (\ref{ber}) specifies a natural map  
of the functor ${\bf R}^{p|q}$ into the functor ${\bf R}^{1|0}$. Analogously, an odd superfield can be considered 
as a natural transformation of functors ${\bf R}^{p|q}\rightarrow {\bf R}^{0|1}$. An arbitrary superfield can be viewed 
as a natural transformation of functors ${\bf R}^{p|q}\rightarrow {\bf R}^{1|1}$ (recall that 
${\bf R}^{1|1}_{\Lambda}=\Lambda_{0}\oplus \Lambda_{1}$ ). Note that  in the above considerations the 
Grassmann algebra can be replaced by any algebra $\Lambda$ every element of which can be represented 
as a sum of a real number and nilpotent element ( assuming that $\Lambda$ is associative, $Z_{2}$-graded , 
supercommutative algebra having unit element). Algebras of this kind will be called almost nilpotent 
algebras or AN algebras. In other words we can say that ${\bf R}^{p|q}$ can be considered as a functor 
on the category of AN algebras and a superfield of definite parity determines a natural transformation 
of the functor  ${\bf R}^{p|q}$ into the functor ${\bf R}^{1|0}$ or ${\bf R}^{0|1}$.

The main notions of superalgebra and of supergeometry can be formulated very easily in the language 
of functors. A superspace can be defined as an arbitrary functor on the category $\bf AN$ of AN 
algebras taking values in the category of sets. The body of a superspace can be defined as the set 
 corresponding to the AN algebra $\Lambda={\bf R}$.  We introduce  a $(p|q)$-dimensional supermanifold 
as a superspace that is locally equivalent to $ {\bf R}^{p|q}$. In other words $(p|q)$-dimensional supermanifold 
can be pasted together from domains in   ${\bf R}^{p|q}$ by means of smooth transformations. 
A body of a $(p|q)$-dimensional supermanifold is a $p$-dimensional smooth manifold. Replacing in the 
definition of a superspace the category of sets by the category of groups or by the category of Lie algebras
we obtain the definitions of supergroup and super Lie algebra respectively.

 It is convenient 
to generalize the notion of the superspace   ${\bf R}^{p|q}$ as follows. 
Denote by  ${\bf R}^{k\oplus l|q}_{\Lambda}$ 
 a set of rows $(x^{1}, \dots, x^{k}, y^{1}, \dots , y^{l}, \xi^{1}, \dots, \xi^{q})$ where $x^{1}, \dots, x^{k}$ 
are arbitrary even elements of AN algebra $\Lambda$ , $y^{1}, \dots , y^{l}$ are nilpotent even elements 
of $\Lambda$ and $\xi^{1}, \dots, \xi^{q}$ are odd elements of $\Lambda$. 
A superspace ${\bf R}^{k\oplus l|q}$ can be defined as a functor $\Lambda \rightarrow {\bf R}^{k\oplus l|q}_{\Lambda}$. 
Strictly speaking to define a functor we should also construct a homomorphism 
$\tilde \alpha: {\bf R}^{k\oplus l|q}_{\Lambda}\rightarrow {\bf R}^{k\oplus l|q}_{\Lambda'}$ for 
every parity preserving homomorphism $\alpha: \Lambda \rightarrow \Lambda'$ of AN algebras. 
We omit this obvious construction.  

Infinite-dimensional analogs of the superspace  ${\bf R}^{k\oplus l|q}$ (partially formal vector superspaces)  are defined in Appendix A.

Let us define a superfield 
on ${\bf R}^{k+l|q}$ as an expression of the form 
\begin{equation}
\sum_{s}\sum_{1\le i_{1} < \dots < i_{s}}
f_{i_{1},\dots ,i_{s}}(x^{1},\dots , x^{k},y^{1},\dots , y^{l})\xi^{i_{1}}\dots \xi^{i_{s}}
\label{3dsuperfield}
\end{equation} 
 where $f_{i_{1},\dots ,i_{s}}$ are smooth functions of variables $(x^{1}, \dots, x^{k})\in {\bf R}^{k}$
 and formal power series with respect to $y^{1}, \dots , y^{l}$. Such an expression determines a map 
of ${\bf R}^{k\oplus l|q} $ into ${\bf R}^{1|0}$ if expression (\ref{3dsuperfield}) is even and into ${\bf R}^{0|1}$ 
if it is odd.

We define a  finite-dimensional partially formal supermanifold (=$(k\oplus l|q)$-dimensional supermanifold) as a superspace that is locally equivalent to 
${\bf R}^{k\oplus l|q} $. 
Almost all notions of (super)geometry 
can be generalized to the case of    $(k\oplus l|q)$-dimensional supermanifolds.  In particular, one can define 
the supergroup of transformations of such a manifold and the corresponding Lie (super) algebra of 
vector fields. As usual, a vector field on a   $(k\oplus l|q)$-dimensional superdomain with coordinates 
$(x^{1}, \dots, x^{k}, y^{1}, \dots , y^{l}, \xi^{1}, \dots, \xi^{q})$ can be identified with a first order 
differential operator 
\begin{equation}
A^{i}\frac{\partial}{\partial x^{i}} + B^{j}\frac{\partial}{\partial y^{j}} + C^{s}\frac{\partial}{\partial \xi^{s}}
\label{deriv}
\end{equation} 
where $A^{i}, B^{j}, C^{s}$ are smooth functions of variables $x^{1}, \dots, x^{k}$ , formal power series 
with respect to $y^{1}, \dots , y^{l}$ and polynomials in $\xi^{1}, \dots, \xi^{q}$. If $A^{i}, B^{j}$ are odd 
and $C^{s}$ are even then the operator (\ref{deriv}) is parity reversing ; we say that the corresponding 
vector field is odd. The definition of an even vector field is similar. 
A $(k\oplus l|q)$-dimensional supermanifold is pasted together from $(k\oplus l|q)$-dimensional superdomains 
by means of smooth transformations. 
If we have a $(p|q)$-dimensional supermanifold $\cal A$ and a $k$-dimensional submanifold ${\cal B}^{(0)}$ 
of its body ${\cal A}^{(0)}$  we can construct easily a $(k\oplus (p-k) |q)$-dimensional supermanifold in 
the following way. For each $\Lambda$ there is a natural mapping 
$\tilde m_{\Lambda}: {\cal A}_{\Lambda} \rightarrow {\cal A}^{(0)}$ 
corresponding to the homomorphism $m_{\Lambda}:\Lambda \rightarrow {\bf R}$ that evaluates a 
numerical part of an element from $\Lambda$. Then, ${\cal B}_{\Lambda}$ consists of those elements 
of ${\cal A}_{\Lambda}$ that project onto ${\cal B}^{(0)}$ under $\tilde m_{\Lambda}$. The mapping 
$\tilde \alpha: {\cal B}_{\Lambda}\rightarrow {\cal B}_{\Lambda'}$ assigned to a homomorphisms 
$\alpha:\Lambda \rightarrow \Lambda'$ is a restriction of the corresponding map defined for the 
supermanifold $\cal A$.  Let us call the supermanifold $\cal B$ obtained this way the restriction  
of supermanifold $\cal A$ to the subset ${\cal B}_{0}$ of its body.

One can define classes of supermanifolds with interesting 
geometric properties restricting the allowed class of coordinate transformations. For example, one can define 
a complex analytic transformation of a $(2k\oplus 2l | 2q)$-dimensional superdomain generalizing the usual
requirement that the Jacobian matrix of transformation commutes with a standard matrix $J$ obeying 
$J^{2}=-1$. Then, a $(k\oplus l|q)$-dimensional complex manifold can be defined as a manifold glued together 
by complex analytic transformations. One can also use complex coordinates 
$(Z^{A})=(z^{1}, \dots , z^{k},w^{1},\dots ,w^{l}, \vartheta^{1}, \dots, \vartheta^{q})$ on a complex manifold. 
They take values in complex AN algebras with antilinear involution.  Of course together with these 
coordinates we should consider complex conjugate coordinates $\bar Z^{A}$. Analytic transformations 
do not mix $  Z^{A}$ and $\bar Z^{A}$. Therefore usually we will not mention  $\bar Z^{A}$. 


Let $U$ be a $(1|N)$-dimensional complex superdomain with complex coordinates 
$(z,\theta^{1},    \dots ,   \theta^{N})$. We  define N-superconformal transformations 
as complex analytic transformations preserving up to a factor the one-form 
$\omega_{N}=dz + \theta^{1}d\theta^{1} + \dots + \theta^{N}d\theta^{N}$ (up to a factor  means here 
up to multiplication by a nonvanishing superfield).
 For example, in case $N=1$ superconformal transformations can be written  explicitly as follows
\begin{eqnarray}
\tilde z&=&u(z)-u'(z)\epsilon (z)\theta \nonumber\\
\tilde \theta &=&\sqrt{u'(z)}\left( \theta + \epsilon (z) +\frac{1}{2}\epsilon (z)\epsilon' (z)\theta \right) 
\label{N=1}   
\end{eqnarray}
where $u(z)$ and $\epsilon (z)$ are even and odd analytic functions of $z$ respectively . 
By a N-superconformal 
manifold we mean a manifold pasted together from $(1|N)$-dimensional complex superdomains by 
N-superconformal transformations. Note that strictly speaking we are looking here not at a 
single supermanifold but rather at a family of supermanifolds parameterized by gluing 
functions similar to $u(z)$ and $\epsilon(z)$ in $N=1$ case. 
The superspace $M_{N}$ of equivalence classes of N-superconformal manifolds is called a moduli space 
of N-superconformal structures (supermoduli space). 

An N-superconformal vector field defined in a complex superdomain $U^{1|N}$ is a vector field $X$ 
such that the Lie derivative of the form $\omega_{N}$ restricted to   $U^{1|N}$ with respect to $X$ 
is proportional to $\omega_{N}$, i.e. $L_{X} \omega_{N} = f\omega_{N}$ for some superfield $f$. Such 
vector fields correspond to infinitesimal N-superconformal transformations. 
Given an N-superconformal compact manifold $\cal A$ one can define an N-superconformal vector field on it 
as a vector field such that its restriction to each elementary coordinate patch $U^{1|N}$ is N-superconformal. 
It follows from  standard results of  
deformation theory that a formal tangent space to the moduli space of N-superconformal structures 
at a ``point" $\cal A$ (being an N-superconformal manifold)  is isomorphic to 
$H^{1}({\cal A}, \gamma_{N})$. Here   $H^{1}({\cal A}, \gamma_{N})$ stands for the first cohomology 
group of $\cal A$ with coefficients in the sheaf of N-superconformal vector fields $\gamma_{N}$.  


Moduli spaces play the central role in the Segal's axiomatics of conformal field theory (CFT) 
( \cite{Segal}). In this 
approach one considers conformal 2d  surfaces (complex curves) with parametrized boundary components. 
Each boundary component is homeomorphic to a circle and it is assumed that the parametrization 
can be extended to a complex coordinate in a small neighborhood.  
We can think of a standard  annulus in the complex plane $\{ z\in {\bf C} | \frac{1}{2} \leq |z| \leq 1 \}$
 mapped 
by a biholomorphic mapping  
into a   neighborhood of the boundary component. The neighborhoods of different boundary components 
are assumed to be non-overlapping. The annuli are divided into two classes: 
``incoming" and ``outgoing". The moduli space of such objects with $m$ incoming and $n$ outgoing 
annuli is denoted by $P_{m,n}$. Let us stress here that we allow disconnected surfaces as well. 
There arise naturally two operations on the sets $P_{m,n}$: 
\begin{eqnarray} \label{disjun} 
P_{m_{1},n_{1}}\times P_{m_{2},n_{2}}& \rightarrow &P_{m_{1}+n_{1}, m_{2} + n_{2}} \\
 P_{m,n} &\rightarrow& P_{m-1,n-1}  \label{sewing}
\end{eqnarray}
Here the first operation corresponds to the disjoint union of surfaces and the second one corresponds
 to the identification of $m$-th incoming annulus with the $n$-th outgoing one by 
 the rule:  $z'=\frac{1}{4}z^{-1}$.  
In Segal's axiomatics CFT is specified by maps
 $$ \alpha_{m,n}: P_{m,n} \rightarrow Hom_{\bf C}\{ H^{m}, H^{n} \} $$ 
assigning to each point in   $P_{m,n}$ a linear mapping $H^{m} \rightarrow H^{n}$ belonging to the trace class
(here $H$ is a fixed Hilbert space). 
The collection of mappings $\alpha_{m,n}$ should satisfy some set of axioms 
ensuring the compatibility of mappings $\alpha_{m,n}$ with mappings (\ref{disjun}), (\ref{sewing}) 
and with permutations of boundary components. 
One can generalize  Segal's axiomatics to the case of N-superconformal field theories replacing 
2d conformal surfaces by N-superconformal manifolds with boundaries and complex annuli by 
N-superconformal annuli (see \cite{AS1} for this type of axiomatics stated for N=2 SCFT).

 
Now let us describe supermanifolds that appear in 2D superconformal field theories (SCFT) having 
different number of supersymmetries for left movers and right movers. 
Consider a $(2|p + q)$-dimensional complex domain $ U$ with even coordinates 
$z_{L}, z_{R}$ and odd coordinates $ \theta_{L}^{1},\dots ,\theta_{L}^{p}, \theta_{R}^{1}, \dots ,\theta_{R}^{q}$. 
The superdomain $U$ can be considered as a superdomain in real superspace ${\bf R}^{4|2p+2q}$. We will 
single out a  subspace $V$ of $U$ that has  real dimension 
$(2\oplus 2|2p+2q)$ by imposing the condition that $z_{R}-\bar z_{L}$ 
is nilpotent. By definition a transformation of $V$ is called $(p,q)$-superconformal if it does not mix the
left coordinates $z_{L}, \theta_{L}^{i}$ with the right coordinates $z_{R},\theta_{R}^{i}$ and preserves up to 
a factor one-forms 
\begin{eqnarray*}
\omega_{L}&=&dz_{L} + \theta_{L}^{1}d\theta_{L}^{1} + \dots + \theta^{p}_{L}d\theta_{L}^{p}\\
\omega_{R}&=&dz_{R} + \theta_{R}^{1}d\theta_{R}^{1} + \dots + \theta^{q}_{R}d\theta_{R}^{q}
\end{eqnarray*} 
We define a $(p,q)$-superconformal manifold as a superspace pasted together from several copies of 
$V$ by means of $(p,q)$-superconformal transformations. Again we assume here that the odd gluing 
parameters such as $\epsilon (z)$ in (\ref{N=1}) are allowed. 


Let ${\cal A}^{(0)}$ be a compact connected oriented 2-dimensional surface of genus $g>1$. 
It is easy to define a moduli space    ${\cal M}_{p,q,g}$ of $(p,q)$-superconformal manifolds having 
${\cal A}^{(0)}$ as a body. Denote by  ${\cal M}_{p,q,g}^{(0)}$ the body of  moduli space 
  ${\cal M}_{p,q,g}$.  One can prove that the superspace ${\cal M}_{p,q,g}$ can be constructed out 
of supermoduli spaces ${\cal M}_{p,g}$ and ${\cal M}_{q,g}$ 
of $p$ and $q$-superconformal manifolds with body ${\cal A}^{(0)}$. 
The construction goes as follows. As in the case of moduli space of conformal structures
${\cal M}_{p,g}$ is equipped with a canonical complex structure. Thus we may consider the superspace
${\cal M}_{p,g} \times \bar {\cal M}_{q,g}$ where $\bar {\cal M}_{q,g}$ denotes the space ${\cal M}_{q,g}$ 
with  conjugate complex structure. Note that there is a natural projection  
$\pi_{p}: {\cal M}_{p,g} \rightarrow {\cal M}_{g}$ to the moduli 
space of complex structures on ${\cal A}^{(0)}$.
 This projection simply corresponds to the fact that each p-superconformal 
manifold has a complex structure on its body. If $z$ is a complex coordinate on ${\cal M}_{g}$ 
and $\bar z$ is its counterpart on $\bar {\cal M}_{g}$ then we define the subset 
${\cal M}_{p,q,g}^{(0)} \subset {\cal M}_{p,g}^{(0)} \times \bar {\cal M}_{q,g}^{(0)}$ as the set of points 
$(a,b)$ such that $\pi_{p}(a)=(\pi_{q}(b))^{*}$. Here $*$ stands for the conjugated complex structure. 
 The superspace  ${\cal M}_{p,q,g}$ has the body 
${\cal M}_{p,q,g}^{(0)}$ and can be obtained as a restriction of ${\cal M}_{p,g} \times \bar {\cal M}_{q,g}$ 
to the corresponding subset of its body. 
Choosing a local coordinate system on ${\cal M}_{p,g} \times \bar {\cal M}_{q,g}$ in such a way that $z, \bar z$ 
are part of the coordinates we see that the condition above implies that $z-(\bar z)^{*}$ can take only 
nilpotent values (as this is zero on the body).   
This means that if ${\cal M}_{p,g}$ is of dimension $(k|l)$ and ${\cal M}_{q,g}$ is of dimension 
$(\tilde k|\tilde l)$ the superspace  ${\cal M}_{p,q,g}$ has the dimension 
$((k+\tilde k -3g+3) \oplus (3g-3) |l+\tilde l)$. 
The construction of  ${\cal M}_{p,q,g}$ presented above is simply a more formal way to say that 
left and right superconformal structures on a $(p,q)$-superconformal manifold are independent 
up to a complex structure on the body that they share. It is easy to generalize this construction to the case 
when ${\cal A}_{0}$ is disconnected.

Now we are in a position to generalize   Segal's axiomatics to include 
$(p,q)$-superconformal field theories. To define a proper analog of the space $P_{m,n}$ 
one should  consider  a larger moduli space of 
(not necessarily connected)
$(p,q)$-superconformal surfaces having  
parameterized boundary components of different type (determined by the boundary conditions for odd 
coordinates). 
Furthermore, the corresponding analogs of the mappings $\alpha_{m,n}$ should
  be holomorphic.  The last requirement sets a connection between $p$-superconformal and 
$(p,p)$-superconformal theories as outlined below. 
 Firstly, it is possible to embed the supermoduli space ${\cal M}_{p,g}$  into ${\cal M}_{p,p,g}$.
Then , given a real analytic function on ${\cal M}_{p,g}\subset {\cal M}_{p,p,g}$
 one can extend it to a holomorphic function defined in some neighborhood of  ${\cal M}_{p,g}$ in 
${\cal M}_{p,p,g}$. Assuming that a continuation to the whole ${\cal M}_{p,p,g}$ is possible 
we see that  a $p$-superconformal field theory determines a $(p,p)$-superconformal one.


Now let us make some remarks about the integration over 
$ {\cal M}_{p,q,g}$. This question is important in particular in heterotic string theory. To calculate a string 
amplitude one defines a holomorphic volume element on $ {\cal M}_{p,q,g}\times \bar {\cal M}_{p,q,g}$ and 
chooses a real cycle of integration on the body of this space. We would like to stress that possible 
reality conditions 
for  odd and nilpotent variables do not affect the integration result (see \cite{KS} or \cite{PD} for discussion ). 
Simply  an integration over odd variables 
is essentially an algebraic operation . 
As for the choice of real cycle on the body of $ {\cal M}_{p,q,g}\times \bar {\cal M}_{p,q,g}$ one can take the 
 (real) diagonal of the body of  $ {\cal M}_{p,q,g}\times \bar {\cal M}_{p,q,g}$. The change of nilpotent variable 
roughly speaking corresponds to the infinitesimal change of integration cycle and therefore does 
not change the  value of integral.


\section*{Appendix A. Superspaces}

In this appendix we develop the approach to definition of superspace in terms of 
functors (see \cite{AS2}).

{\it The notion of superspace. } 
Let us introduce a notion of almost nilpotent algebra (or  AN algebra ).
An AN algebra is an associative finite dimensional $Z_{2}$-graded supercommutative 
algebra $A$ with unit element such that its ideal of nilpotent elements has codimension 1.
In other words $A$ can be decomposed  (as a vector space)  in a direct sum $A=R+N$ 
where $R$ is the canonically embedded  ground field (  {\bf R} or  {\bf C} ) and  $N = N(A)$ consists of nilpotent elements. 
An example of AN algebra is a Grassmann algebra. Another important example is an 
algebra generated by unit element $1$ and an element $x$ satisfying the only relation $x^n=0$.
An example including the previous two is   a supercommutative algebra  with 
odd generators $\xi_{1}, \xi_{2}, \cdots , \xi_{n}$ and even generators $ x_{1}, x_{2}, \cdots ,x_{m}$
satisfying the relations $x_{1}^{n_{1}}=0 , x_{2}^{n_{2}}=0, \cdots, x_{m}^{n_{m}}=0$. 
Below parity preserving homomorphisms of AN algebras are called  morphisms .  

Now we define a superspace $S$ as a rule  assigning  to each AN algebra $\Lambda$ a set 
$S_{\Lambda}$ that we call the set of $\Lambda$-points of $S$ and to each morphism of two 
AN algebras $\alpha: \Lambda \rightarrow \Lambda'$ a map $\tilde \alpha:S_{ \Lambda }
\rightarrow S_{ \Lambda'}$ of the corresponding sets in a  way that is consistent with compositions
 of morphisms. The last assertion means that the map $\widetilde{\alpha_{2}\alpha_{1}}$
corresponding to the composition of 
morphisms $\alpha_{2}\alpha_{1}$  is equal to the composition of maps $\tilde \alpha_{2}\tilde \alpha_{1}$.
We consider all AN algebras together with parity preserving homomorphisms between them 
 as a category $\bf AN$.  
In  standard mathematical terminology superspace is a covariant functor from the category $\bf AN$ to 
the category of sets. 

Let us give a  definition of  a mapping of one superspace into another. Let $\cal N$ and $\cal M$ be 
superspaces. We say that there is a mapping $F: {\cal N} \rightarrow  {\cal M}$ if for every AN algebra 
$\Lambda$ there is a map $F_{\Lambda}: {\cal N}_{\Lambda} \rightarrow  {\cal M}_{\Lambda}$ and
  the maps $F_{\Lambda}$ are consistent with maps $\tilde \alpha$ corresponding to the 
morphisms of AN algebras $\alpha: \Lambda \rightarrow \Lambda'$. In other words  diagrams of the following 
type are commutative

\begin{eqnarray} 
 {\cal N}_{\Lambda}&\smash{ 
 \mathop{\longrightarrow}\limits^{F_{\Lambda}}}
& {\cal M}_{\Lambda} \nonumber \\ 
 \tilde \alpha \Big\downarrow& &\Big\downarrow \tilde \alpha \label{diagram}\\
 {\cal N}_{\Lambda'}&\smash{ 
 \mathop{\longrightarrow}\limits^{F_{\Lambda'}}}
 &{\cal M}_{\Lambda'} \nonumber\\  \nonumber
\end{eqnarray}
In  standard terminology a mapping of superspaces is a natural transformation of functors. 
Isomorphism of superspaces is defined as an isomorphism of functors.

The space $S_{R}$ corresponding to an AN algebra  R (i.e. an AN algebra having a trivial nilpotent part N) 
is called the body of  superspace $S$. For every AN algebra $\Lambda$ there is a homomorphism 
$m: \Lambda \rightarrow  R$ assigning to an element of $\Lambda$ its projection onto R with 
respect to the canonical decomposition $\Lambda={ R+N(\Lambda)}$. The corresponding  map 
$\tilde m: S_{\Lambda} \rightarrow S_{R}$ associates with each $\Lambda$-point $a \in S_{\Lambda}$ 
a point $\tilde m(a)$ in the body of $S$ that we call the numeric part of $a$. From now on if not specified 
we will assume our ground field is the field of real numbers $\bf R$. Here we give some basic 
examples of superspaces.

{\bf Example 1} Let $V$ be a $Z_{2}$ graded vector space, i.e. $V=V_{0} \oplus V_{1}$ where $V_{0}$ and 
$V_{1}$ are called the even and odd subspaces respectively. We define the set $V_{\Lambda}$ to be 
the set of all formal (finite) linear combinations $\sum_{i} e_{i}\alpha_{i}   +   \sum_{j} f_{j}\beta_{j} $ where 
$e_{i}\in V_{0}, f_{j}\in V_{1}$ , $\alpha_{i}$ and $\beta_{j}$ are respectively arbitrary even and odd 
elements from $\Lambda$ (we assume the following identifications 
$(\alpha_{1}+\alpha_{2} )v=\alpha_{1}v + \alpha_{2}v$ ,  
$\alpha (v_{1} + v_{2}) = \alpha v_{1}  + \alpha v_{2} $ where $\alpha, \alpha_{1}, \alpha_{2} \in \Lambda$ 
and $v, v_{1}, v_{2} \in V$ ). The mapping $\tilde \rho$ corresponding to a morphism 
$\rho : \Lambda \rightarrow \Lambda'$ carries the point  
$\sum_{i} e_{i}\alpha_{i}   +   \sum_{j} f_{j}\beta_{j} $ to $\sum_{i} e_{i}\rho(\alpha_{i})   +   
\sum_{j} f_{j}\rho(\beta_{j}) $ that obviously lies in $V_{\Lambda'}$. 
One can easily  see that the maps $\tilde \rho $ behave properly under compositions. 
Note that each set $V_{\Lambda}$ is a linear 
space itself and is equipped with a canonical structure of (left or right) $\Lambda_{0}$ module where $\Lambda_{0}$ 
is an even part of $\Lambda$. For the last statement we set 
$\mu (\sum_{i} e_{i}\alpha_{i}   +   \sum_{j} f_{j}\beta_{j} ) = 
\sum_{i} e_{i}(\mu\alpha_{i})   +   \sum_{j} f_{j}(\mu\beta_{j})$ for any $\mu \in \Lambda_{0}$ 
(this is for the left $\Lambda_{0}$ module 
structure, the modification for the   structure  of right $\Lambda_{0}$ module is obvious). 
 We call superspaces 
with this property linear superspaces. The construction given above works for any $Z_{2}$-graded linear 
space, including  infinite-dimensional ones. In the  finite-dimensional case the corresponding 
superspace is called $(p|q)$-dimensional linear 
superspace and is denoted by ${\bf R}^{p|q}$ (here $p=dim V_{0}, q=dim V_{1}$).
 Choosing   a basis in $V_{0}$ and $V_{1}$ one obtains 
a representation of $ {\bf R}^{p|q}_{\Lambda}$ as a set of $(p+q)$-tuples of elements from $\Lambda$ 
where first $p$ elements are from $\Lambda_{0}$ and the next $q$ elements are from $\Lambda_{1}$.  
It readily follows that the body of ${\bf R}^{p|q}$ is naturally isomorphic (as a linear space) to ${\bf R}^{p}$.

In the infinite-dimensional case, we will use the notation ${\bf R} (V_0,V_1))$ for a superspace corresponding to a $\bf{Z}_2$-graded vector space $V_0+V_1$.

{\bf Example 2} Here we want to describe some general operations over superspaces. Let $\cal M$ be 
a superspace with a body ${\cal M}^{(0)}\equiv {\cal M}_{\bf R}$. Let ${\cal N}^{(0)} \subset {\cal M}^{(0)}$ 
be any subset. Define ${\cal N}_{\Lambda} = (\tilde m ({\cal N}^{(0)}) )^{-1} \subset {\cal M}_{\Lambda}$ 
where $\tilde m$ is the described above map evaluating the numerical part of $\Lambda$-point. 
For any morphism of AN algebras $\rho : \Lambda \rightarrow \Lambda'$  we can define a map 
$ {\cal N}_{\Lambda} \rightarrow   {\cal N}_{\Lambda'}$ as the restriction of $\tilde \rho$ to 
${\cal N}_{\Lambda}$. Then it is easy to check that the sets ${\cal N}_{\Lambda}$ together with maps 
$\tilde \rho$ form a superspace. This new superspace has the set ${\cal N}^{(0)}$ as its body. 
The superspace constructed this way out of the given superspace $\cal M$ and a subset  
 ${\cal N}^{(0)} \subset {\cal M}^{(0)}$  will be called the restriction of superspace $\cal M$ to the set 
 ${\cal N}^{(0)}$. Given a domain ${\rm U} \subset {{\bf R}^{p}}$ we can construct a restriction of superspace 
${\bf R}^{p|q}$ to $\rm U$ which is called a $(p|q)$-dimensional superdomain and denoted as $\rm U^{p|q}$. 

Similarly, if we assume that $V_0$ is a topological vector space we can define a superdomain ${\bf U}(V_0,V_1)$ in $\bf{R}(V_0,V_1)$ for every open set  $U\subset V_0.$

{\bf Example 3} Given a linear subspace ${\bf R}^{m} \subset {\bf R}^{p}$ one can construct the 
restriction of superspace  ${\bf R}^{p|q}$ to this subset. We denote this new superspace by 
${\bf R}^{m\oplus n|q}$ where $n=p-m$. The explicit construction of it is as follows. 
Let $V=V_{0} \oplus V_{1}$ be a $Z_{2}$ graded vector space  and 
$dim V_{0}=p, dim V_{1}=s$. Given a decomposition $V_{0}=U_{0} \oplus W_{0}$ 
($dim U_{0}=m, dim W_{0}=n$)  we  
define $ {\bf R}^{m\oplus n|s}_{\Lambda}$ to be the set of  linear combinations 
of the form $\sum_{i}d_{i}\alpha_{i} + \sum_{j}e_{j}\beta_{j} + \sum_{k}f_{k}\gamma_{k}$ where 
$d_{i}\in U_{0}, e_{j}\in W_{0}, f_{k}\in V_{1}$ and $\alpha_{i}\in \Lambda_{0}, \beta_{j}\in 
\Lambda_{0}\cap N(\Lambda),  \gamma_{k}\in \Lambda_{1}$
(as above we require the natural distributivity conditions). Given bases of $U_{0}, W_{0}, V_{1}$ 
we can represent a point from  $ {\bf R}^{m\oplus n|s}_{\Lambda}$ as a   $p+q$-tuple of elements from $\Lambda$ 
in which first m elements are even, next n elements are even and nilpotent and the last s elements are odd. 
Similarly to the construction of ${\bf R}^{p|q}$ for any morphism of AN algebras $\rho$ we can construct 
a mapping $\tilde \rho$  that carries the point 
$\sum_{i}d_{i}\alpha_{i} + \sum_{j}e_{j}\beta_{j} + \sum_{k}f_{k}\gamma_{k}$ to 
 $\sum_{i}d_{i}\rho(\alpha_{i}) + \sum_{j}e_{j}\rho(\beta_{j}) + \sum_{k}f_{k}\rho(\gamma_{k})$.  
Given a domain $\rm U \subset  R^{m}$ the restriction of ${\bf R}^{m\oplus n|s}$ to $\rm U$ is called an
$(m\oplus n|s)$-dimensional superdomain and is denoted by $\rm U^{m\oplus n|s}$. 
Note that like ${\bf R}^{p|q}$, ${\bf R}^{m\oplus n|q}$ is a vector superspace.

Notice that this construction has an obvious infinite-dimensional generalization.
Let us consider a triple $M,N,S$ of topological linear vector spaces and a superspace
 ${\bf R}(M+N,S)$ corresponding to a ${\bf Z}_2$-graded space $V_0+V_1$ where $V_0=M+N, V_1=S.$ Then we can define  a partially formal linear superspace restricting 
 ${\bf R}(M+N,S)$ to $M.$ Fixing an open subset $U\subset M$ we can define a partially formal superdomain ${\bf  U}(M+N,S)$

In a more interesting situations we have a superspace $\cal M$ such that  the sets ${\cal M}_{\Lambda}$  are 
equipped with some additional structure and the maps $\tilde \rho$ are consistent with it. 
In other words we can consider a functor from the category $\bf AN$ to a category 
of sets with  additional structure , e.g. smooth manifolds, groups, Lie groups, Lie algebras, etc. 
In the previous section we gave an example of this sort of object, namely a vector superspace is 
a functor with values in linear spaces.

 Similarly, one can define a topological superspace as a 
functor from the category {\bf AN} to the category of topological spaces.

Next we introduce the notions of associative superalgebra, Lie superalgebra and  supergroup. 
An associative superalgebra  is a functor $\cal A$ taking values in the category of associative 
algebras. In addition we require that each algebra ${\cal A}(\Lambda )$ is equipped with a 
structure of $\Lambda_{0}$-module in a way that is consistent with associative multiplication 
and the mappings $\tilde \alpha$. 
 An example of an associative superalgebra can be obtained in the following way. 
Given a $Z_{2}$-graded associative algebra $A$ one can consider  linear combinations 
whose coefficients are elements of AN algebra $\Lambda$ with appropriate parity (as in the construction 
of  vector superspace).  
Setting $(a\lambda)(b\mu)=\pm (ab)(\lambda\mu)$ for all $(a,b) \in A, \quad 
\lambda, \mu \in \Lambda$ (where minus sign occurs only if $\lambda$ and $b$ are both odd) we get an associative 
algebra associated with every $A$ and  AN algebra $\Lambda$. 
These associative algebras can be considered as algebras on sets of $\Lambda$-points of a certain superspace. 
A Lie superalgebra is a functor with values in Lie algebras equipped with a natural structure of linear 
superspace. Analogously to the previous example one can construct a Lie superalgebra out of any $Z_{2}$-graded 
Lie algebra. Another example is a Lie superalgebra constructed out of an associative superalgebra. 
One simply defines a Lie algebra structure on the sets of $\Lambda$-points via the commutator.

 A supergroup is a functor with values in the category of groups. 
An important example is a supergroup $GL(m|n)$ whose set of $\Lambda$-points consists 
of $(m+n) \times (m+n)$ nondegenerate 
block matrices with 
entries from $\Lambda$. 
The entries of $m \times m$ and $n \times n$ blocks are even and the entries of 
$m \times n$ and $n \times m$ 
ones are odd. A matrix is called nondegenerate if the left inverse matrix exists (one can show that this implies 
the existence of right inverse matrix ). The set of all such block matrices (not necessarily nondegenerate) is 
an example of associative superalgebra. 


We say that a supergroup $G$ acts on a superspace $\cal S$ if for any AN algebra $\Lambda$ 
the group $G_{\Lambda}$ acts on the set ${\cal S}_{\Lambda}$ in a way that is consistent with 
mappings $\tilde \alpha$ between $\Lambda$-points. Namely, if $\phi_{g}$ is the mapping of the set 
${\cal S}_{\Lambda}$ to itself corresponding to the  element $g\in G_{\Lambda}$, then 
$\tilde \alpha \circ \phi_{g} = \phi_{\tilde \alpha (g)} \circ \tilde \alpha$. The collection of quotient  
spaces ${\cal S}_{\Lambda} / G_{\Lambda}$ is endowed with the structure of superspace   in a natural way.


We say that a topological superspace $\cal M$  is a partially formal supermanifold if it is pasted together from partially formal superdomains. In other words, we assume that the body  ${\cal M}_{\bf R}$ of this superspace can be covered by open sets $U$ in such a way that for every $U$  the restriction of $\cal M$ to $U$ is isomorphic to a partially formal domain as a topological superspace.

\smallskip

\section* {Appendix B. Smooth supermanifolds.}
 We define   
a smooth superspace $\cal M$ as a functor with values in smooth manifolds satisfying some 
additional conditions to be described in a moment.  Recall that in the 
examples described above spaces of $\Lambda$-points of linear superspaces were equipped with 
a structure of $\Lambda_{0}$-module. Thus we will require that a tangent space at each point in 
${\cal M}_{\Lambda}$ is equipped with a structure of (left) $\Lambda_{0}$-module.  
A mapping $\tilde \alpha:M_{\Lambda} \rightarrow M_{\Lambda'}$ corresponding to a morphism $\alpha$ 
of two AN algebras is called $\Lambda_{0}$-smooth if the corresponding 
tangent map $(\tilde \alpha)_{*}$ commutes with $\Lambda_{0}$-module structures, i.e. 
 $(\tilde \alpha)_{*}\circ \lambda_{0} = \alpha (\lambda_{0}) \circ (\tilde \alpha)_{*}$.

In this section we would like to give a definition of a supermanifold as a superspace with values in the 
category of smooth manifolds  which is glued together from $(m\oplus n|s)$-dimensional superdomains. 
To this end we need to describe an appropriate class of gluing transformations.

Our next goal will be to describe all possible mappings $F: {\bf R}^{p\oplus q|s} \rightarrow  
{\bf R}^{p'\oplus q'|s'}$. 
 Note that for now we do 
not require our maps to be consistent with natural $\Lambda_{0}$-module structure defined on 
${\bf R}^{p\oplus q|s}_{\Lambda}$ and ${\bf R}^{p'\oplus q'|s'}_{\Lambda}$, nor do we impose any smoothness 
conditions on $F_{\Lambda}$. We introduce an algebra $A_{p,q,s}$ consisting of the expressions of 
the form 

\begin{equation} \label{A1}
\omega = \sum_{i=0}^{s} \sum_{1 \le \alpha_{1} < \cdots < \alpha_{i} \le s} f^{\alpha_{1}, \cdots , \alpha_{i}} 
\xi_{\alpha_{1}}\dots \xi_{\alpha_{i}} 
\end{equation} 
where 
\begin{equation}
f^{\alpha_{1}, \cdots , \alpha_{i}} = 
 \sum_{(\beta_{1}, \cdots \beta_{p},\gamma_{1}, \cdots, \gamma_{q})} 
G^{\alpha_{1}, \cdots , \alpha_{i}}_{\beta_{1}, \cdots \beta_{p},\gamma_{1}, \cdots, \gamma_{q}} 
(m_{1}, \cdots, m_{p}) n_{1}^{\beta_{1}}\dots n_{p}^{\beta_{p}}l_{1}^{\gamma_{1}}\dots l_{q}^{\gamma_{q}} 
 \label{A2}
\end{equation}
 are formal series in variables $n_{1}, \cdots, n_{p},l_{1}, \cdots, l_{q}$, the coefficients of which are arbitrary 
functions of  real variables $m_{1}, \cdots, m_{p}$. If one assumes that $\xi_{1}, \cdots ,\xi_{s}$ in 
(\ref{A1}) are generators of Grassmann algebra one can define the parity of expression (\ref{A1}) in a natural way. 
Given an even (odd) element  $\omega\in A_{p,q,s}$ one can construct a mapping  $\alpha_{\Lambda}^{\omega}$ of 
${\bf R}^{p+q|s}_{\Lambda}$ into ${\bf R}^{1|0}_{\Lambda}$ (respectively, into   ${\bf R}^{0|1}_{\Lambda}$) in 
the following way. Given a $\Lambda$-point 
$X=(x_{1}, \dots , x_{p}, y_{1}, \dots, y_{q}, \theta_{1},\dots , \theta_{s})\in  {{\bf R}^{p+q|s}_{\Lambda}}$  
 to evaluate $\alpha_{\Lambda}^{w}(X)$ one should substitute in (\ref{A1}), (\ref{A2}) odd elements 
$\theta_{i}$ for $\xi_{i}$  , numeric part $m(x_{i})$ for $m_{i}$ , nilpotent part $x_{i}-m(x_{i})$ for  
$n_{i}$ and $y_{i}$ for $l_{i}$. Since $y_{i}$ and $x_{i}-m(x_{i})$ are nilpotent all formal power series 
terminate. 
Given an even element  $\omega \in A_{p,q,s}$ with the vanishing free term one can construct a map 
$\alpha_{\Lambda}^{\omega}: {\bf R}^{p\oplus q|s}_{\Lambda} \rightarrow {\bf R}^{0\oplus 1|0}$ using 
the same substitutions  as those described above.  It can be readily verified that the mappings 
$\alpha_{\Lambda}^{\omega}$ for different $\Lambda$ are consistent, i.e. the corresponing diagrams
of the form (\ref{diagram}) are commutative, and thus define a mapping of superspace ${\bf R}^{p\oplus q|s}$ into the
superspace ${\bf R}^{1|0}$, ${\bf R}^{0|1}$ or ${\bf R}^{0\oplus 1|0}$ if 
$\omega$ is even, odd or even and nilpotent respectively. In the same fashion,  
given a row $L$ consisting of $p'$ even, $q'$ even and nilpotent  and 
$s'$ odd elements from $A_{p,q,s}$ one can construct a map 
$\alpha^{L}_{\Lambda}:  {\bf R}^{p\oplus q|s}_{\Lambda}  \rightarrow {\bf R}^{p'\oplus q'|s'}_{\Lambda}$. The maps 
$\alpha^{L}_{\Lambda}$ for different $\Lambda$ are mutually consistent and determine a map  
$\alpha^{L}$ of the corresponding superspaces. In fact  maps of the form $\alpha^{L}$ exhaust all 
maps between superspaces ${\bf R}^{p\oplus q|s}$ and ${\bf R}^{p'\oplus q'|s'}$. More precisely, we have 


\proclaim Theorem 1. The constructed correspondence of rows of $p'$ even, $q'$ even and nilpotent
 and $s'$ odd elements from $A_{p,q,s}$ to  mappings 
${\bf R}^{p\oplus q|s}\rightarrow  {\bf R}^{p'\oplus q'|s'}$ is a bijection. \par
{\it Proof.} 
The injectivity follows directly from the construction of correspondence. Thus we only need to show that the 
map is onto. Assume the mappings 
$\alpha_{\Lambda}:{\bf R}^{p\oplus q|s}_{\Lambda} \rightarrow  {\bf R}^{p'\oplus q'|s'}_{\Lambda}$ define a mapping 
of corresponding superspaces. Let $\Lambda_{k, r,N}$  be an AN algebra generated by the even 
generators $a_{1}, \dots, a_{k}$ satisfying the relations $a_{1}^{N}=a_{2}^{N}=\dots=a_{k}^{N}=0$ and 
odd generators $\eta_{1}, \dots ,\eta_{r}$. We assume the only relations in   $\Lambda_{k, s,N}$ are  
due to supercommutativity and those  relations on even generators that are written above. 
 Let $\Lambda$ be any AN algebra and 
$x=(m_{1} + n_{1}, \dots, m_{p} + n_{p}, l_{1}, \dots, l_{q}, \theta_{1}, \dots , \theta_{s})$ be any point 
from   ${\bf R}^{p+q|s}_{\Lambda}$ (here $m_{i} + n_{i}$ is the decomposition into numerical and nilpotent parts 
respectively). Denote by  $\tau_{m,N}$ a point 
$(m_{1} + a_{1}, \dots, m_{p} + a_{p}, a_{p+1}, \dots , a_{p+q}, \eta_{1}, \dots, \eta_{s})\in 
{\bf R}^{p\oplus q|s}_{\Lambda_{p+q,s,N}}$ (the lower script $m$  stands for the collection 
$(m_{1}, \dots , m_{p})$) . We claim that if N is sufficiently large there exists a morphism 
$\rho: \Lambda_{p+q,s,N} \rightarrow \Lambda$ such that $\tilde \rho (\tau_{m,N}) = x$. 
Indeed, let $n(\Lambda)$ be the maximal nilpotency degree of elements in $\Lambda$. Then, for 
$N\ge n(\Lambda)$ the  map
 
$$a_{1}\mapsto n_{1}, \dots ,a_{p} \mapsto n_{p}$$ 
$$ a_{p+1} \mapsto l_{1}, \dots , a_{p + q} \mapsto l_{q}$$
$$ \eta_{1} \mapsto \theta_{1}, \dots , \eta_{s} \mapsto \theta_{s}$$ 
can be extended up to a parity preserving homomorphism 
$\rho: \Lambda_{p+q,s,N} \rightarrow \Lambda$. One can easily see that for  $\rho$ constructed
this way we have $\tilde \rho (\tau_{m,N}) = x$. By commutativity of diagram (\ref{diagram}) corresponding to the 
homomorphism $\rho$ we have $\tilde \rho \alpha_ {\Lambda_{p+q,s,N}} =  \alpha_{\Lambda} \tilde \rho$. 
 Therefore $\alpha_{\Lambda}(x) = \alpha_{\Lambda}\tilde \rho (\tau_{m,N}) =  
\tilde \rho \alpha_ {\Lambda_{p+q,s,N}}(\tau_{m,N})$ which means that the image of $x$ under the map 
$\alpha_{\Lambda}$ is uniquely determined by the image of $\tau_{m}$ under the map  
$\alpha_ {\Lambda_{p+q,s,N}}$. Since $\Lambda$ is an arbitrary AN algebra we see that the mapping 
of superspaces $\alpha$ is determined by the images of points 
$\tau_{m,N}\in {\bf R}^{p\oplus q|s}_{\Lambda_{p+q,s,N}}$ for different $N$. 
Since the nilpotent parts of the coordinates 
 of $\tau_{m,N}$ are the generators of $\Lambda_{p+q,s,N}$ the coordinates of the image of $\tau_{m,N}$ under 
$\alpha_{\Lambda_{p+q,s,N}}$ are clearly  polynomials in them with coefficients depending on 
$m_{1},\dots, m_{p}$ and hence define  elements from $A_{p,q,s}$. 
 The last step is to show that the polynomials corresponding to different $N$ are 
consistent in a sense that  polynomials corresponding to algebras with smaller $N$ are just initial pieces of 
those corresponding to the algebras with larger $N$. This is due to the fact that there exists a homomorphism 
$\rho: \Lambda_{p+q,s,M} \rightarrow  \Lambda_{p+q,s,N}$ for $M\ge N$ such that $\tilde \rho(\tau_{m,M}) = 
\tau_{m,N}$. The homomorphism $\rho$ is defined by the following maps 
$$ \tilde a_{i} \mapsto a_{i}, \quad \tilde \eta_{j} \mapsto \eta_{j}$$ 
where $\tilde a_{i}, a_{i}$ are even and $\tilde \eta_{j}, \eta_{j}$ are odd generators of  
$\Lambda_{p+q,s,M}$ and  $\Lambda_{p+q,s,N}$ respectively. The same use of commutative diagram 
(\ref{diagram}) for thus constructed $\rho$  as above assures that the polynomials corresponding to 
$\alpha_{\Lambda_{p+q,s,N}}(\tau_{m})$ are consistent and thus define in a unique way a row of $p'$ even, 
$q'$ even  and nilpotent and $s'$ odd elements from $A_{p,q,s}$. $\Box$


Let us turn now to some additional requirements 
that we impose on admissible mappings of supermanifolds. First note that if $\Lambda$ is an AN 
algebra such that $dim \Lambda_{0}=k, dim \Lambda_{1}=l$ then we have an isomorphism 
${\bf R}^{p\oplus q|s}_{\Lambda} \cong {\bf R}^{L}, \quad L=pk+q(k-1)+sl$. Since the tangent space  
$\rm TR^{L}$ is isomorphic to ${\bf R}^{L}$ itself,  for every $\Lambda$ point 
$x\in {\bf R}^{p\oplus q|s}_{\Lambda}$ we have a $\Lambda_{0}$ module structure induced on $T_{x}$ 
in a natural way. We say that a smooth mapping 
$\alpha_{\Lambda}: {\bf R}^{p\oplus q|s}_{\Lambda} \rightarrow  {\bf R}^{p'\oplus q'|s'}_{\Lambda}$ is 
$\Lambda_{0}$-smooth  if for every  $x\in {\bf R}^{p\oplus q|s}_{\Lambda}$ the tangent map 
$(\alpha_{\Lambda})_{*}:T_{x} \rightarrow T_{\alpha_{\Lambda}(x)}$ is a homomorphism of $\Lambda_{0}$ 
modules. The mapping $\alpha: {\bf R}^{p\oplus q|s} \rightarrow {\bf R}^{p'\oplus q'|s'}$ is said to be smooth if 
for all AN algebras $\Lambda$ the maps $\alpha_{\Lambda}$ are $\Lambda_{0}$-smooth. The description of 
smooth mappings between the spaces   ${\bf R}^{p\oplus q|s}$ is given by theorem 2 which we will formulate 
in a moment. But first we introduce the algebra $B_{p,q,s}$ as a tensor product of 
$C^{\infty}({\bf R}^{p})$, an algebra of formal series with respect to $q$ variables  $n_{1}, \dots , n_{q}$ and 
a Grassmann algebra with $s$ generators  $\xi_{1}, \dots, \xi_{s}$ :  

$$B_{p,q,s}=C^{\infty}({\bf R}^{p})\otimes{\bf R}[[n_{1}, \dots , n_{q}]]\otimes \Lambda^{.}(\xi_{1}, \dots, \xi_{s})$$ 

 More explicitly, elements of  $B_{p,q,s}$ are of  the form (\ref{A1}) where 
coefficients $f^{\alpha_{1}, \cdots , \alpha_{i}}$ are series of the form 
\begin{equation} \label{B} 
f^{\alpha_{1}, \cdots , \alpha_{i}} = 
 \sum_{(\beta_{1}, \cdots \beta_{q})} 
G^{\alpha_{1}, \cdots , \alpha_{i}}_{\beta_{1}, \cdots \beta_{q}} 
(x_{1}, \cdots, x_{p}) l_{1}^{\beta_{1}}\dots l_{p}^{\beta_{q}}
\end{equation} 
and $G^{\alpha_{1}, \cdots , \alpha_{i}}_{\beta_{1}, \cdots \beta_{q}} 
(x_{1}, \cdots, x_{p})$ are smooth functions of variables $x_{1}, \cdots, x_{p}$. 
Again the $Z_{2}$ grading in the 
exterior algebra $\Lambda^{.}(\xi_{1}, \dots, \xi_{s})$ induces a $Z_{2}$ grading on $B_{p,q,s}$. 
One can check that given 
a $\Lambda$-point from ${\bf R}^{p\oplus q|s}$ and  an even (odd) element from $B_{p,q,s}$ by substituting 
first $p$ coordinates instead of variables $x_{1}, \dots, x_{p}$
, next $q$ coordinates instead of variables $l_{1}, \dots , l_{q}$ and the last $s$ variables instead of 
$\xi_{1}, \dots , \xi_{s}$ one gets an even (odd) element from $\Lambda$. These substitutions make sense 
because one can use the Taylor expansion of functions $G$ in the nilpotent parts of the first $p$ variables and 
the formal series in the  next $q$ variables terminate due to  nilpotency. In parallel with the case of algebra 
$A_{p,q,s}$ one can check that a row of $p'+q'+s'$ elements from $B_{p,q,s}$ with 
appropriate parities and nilpotency properties defines a mapping 
${\bf R}^{p\oplus q|s} \rightarrow  {\bf R}^{p'\oplus q'|s'}$. 
 Moreover, this map is smooth in the sense explained above. Note that such a row   also defines a parity 
preserving homomorphism $B_{p',q',s'} \rightarrow B_{p,q,s}$ (the generators  
are sent to the corresponding elements of the row). In fact the following theorem is valid:

  
\proclaim Theorem 2.  The smooth mappings ${\bf R}^{p\oplus q|s}\rightarrow {\bf R}^{p'\oplus q'|s'}$ are in 
a bijective correspondence with  the parity preserving homomorphisms 
$B_{p',q',s'} \rightarrow B_{p,q,s}$ .  \par

{\it Proof.} 
We have shown how to construct a smooth mapping ${\bf R}^{p\oplus q|s}\rightarrow  {\bf R}^{p'\oplus q'|s'}$ 
out of a row of $p'$ even, 
$q'$ even  and nilpotent and $s'$ odd elements from $B_{p,q,s}$. Conversely, given a mapping 
$\alpha: {\bf R}^{p\oplus q|s}\rightarrow  {\bf R}^{p'\oplus q'|s'}$ by theorem 1 we can assign to it a row of 
elements from $A_{p,q,s}$. Let $\omega_{i}=\omega_{i}(m,n,l,\xi)$ (see (\ref{A1}), (\ref{A2}) ) 
be the $i$-th coordinate of this row. From the assumption 
of $\Lambda_{0}$-smoothness it follows directly that the coefficients 
$G^{\alpha_{1}, \cdots , \alpha_{j}}_{\beta_{1}, \cdots \beta_{p},\gamma_{1}, \cdots, \gamma_{q}} 
(m_{1}, \cdots, m_{p})$ corresponding to elements $\omega_{i}$ are smooth functions. 

Let $\lambda=m(\lambda) + n(\lambda)$ be  an element  of $\Lambda_{0}$. 
 Set $ m(\lambda) \epsilon$ and $ n(\lambda) \epsilon$ to be an increment of numerical and nilpotent parts 
of the $k$-th coordinate of the space ${\bf R}^{p\oplus q|s}_{\Lambda}$ respectively 
( $k \le p$, $\epsilon$ is a small real number). 
All other coordinates except the $k$-th one are fixed. 
Then, by $\Lambda_{0}$-smoothness 
$$\omega_{i} (m_{k} + m(\lambda) \epsilon, n_{k} + n(\lambda) \epsilon) - 
\omega_{i} (m_{k} , n_{k}) =  \lambda (\omega(m_{k} + \epsilon, n_{k}) - \omega(m_{k},n_{k})) + o(\epsilon)$$
On the other hand the LHS of the last equation can be written as  
$$m(\lambda ) \epsilon \frac{ \partial \omega_{i}(m_{k}, n_{k})}{\partial m_{k}} + 
n(\lambda) \epsilon  \frac{ \partial \omega_{i}(m_{k}, n_{k})}{\partial n_{k}} + o(\epsilon)$$
from which we conclude that 
$\frac{ \partial \omega_{i}(m_{k}, n_{k})}{\partial n_{k}} = \frac{ \partial \omega_{i}(m_{k}, n_{k})}{\partial m_{k}}$. 
The last equation being written in terms of series $\omega_{i}=b_{0} + b_{1}n_{k} + b_{2}n_{k}^{2} + \dots$ 
where $b_{i}$ depend on $m,l,\xi$ and all $n$'s except $n_{k}$ reads as 
$$
b_{1} + 2b_{2}n_{k} + 3b_{3}n_{k}^{2} + \dots = \frac{\partial b_{0}}{\partial m_{k}} + 
  \frac{\partial b_{1}}{\partial m_{k}}n_{k}  + \frac{\partial b_{2}}{\partial m_{k}}n_{k}^{2} + \dots 
$$ 
Therefore, 
$$ \omega_{i}= b_{0} + \frac{1}{1!}\frac{\partial b_{0}}{\partial m_{k}}n_{k} + 
\frac{1}{2!}\frac{\partial^{2} b_{0}}{\partial m_{k}^{2}}n_{k}^{2} + \dots $$ 
Hence $\omega_{i}=b_{0}(m_{k}+n_{k})$ where for shortness all other arguments are skipped. 
Repeating this argument successively for all $k=1, \dots , p$ we get 
$$ \omega_{j}=\sum_{i=0}^{s} \sum_{1 \le \alpha_{1} < \cdots < \alpha_{i} \le s} f_{j}^{\alpha_{1}, \cdots , \alpha_{i}} 
\xi_{\alpha_{1}}\dots \xi_{\alpha_{i}} $$ 
where 
$$
f_{j}^{\alpha_{1}, \cdots , \alpha_{i}} = 
 \sum_{(\gamma_{1}, \cdots, \gamma_{q})} 
G^{\alpha_{1}, \cdots , \alpha_{i}}_{j; 0, \cdots 0,\gamma_{1}, \cdots, \gamma_{q}} 
(m_{1}+n_{1}, \cdots, m_{p}+n_{p})l_{1}^{\gamma_{1}}\dots l_{q}^{\gamma_{q}} 
$$ 
Compare with (\ref{A1}), (\ref{B}).  
The last two expressions mean precisely that $\omega$ is given by a row of $p'+q'+s'$ elements (with corresponding parity 
and nilpotency properties) from 
the algebra $B_{p,q,s}$, i.e. represents a parity preserving homomorphism $B_{p',q',s'} \rightarrow B_{p,q,s}$. 
$\Box$

The theorems 1 and 2 are modifications of statements proved in \cite{SVor}.

The theorem 2 has a straightforward generalization to the case of mappings between superdomains. 
Let $B_{p,q,s}(U)={\bf R}[[n_{1}, \dots , n_{q}]]\otimes C^{\infty}({U})\otimes \Lambda^{.}(\xi_{1}, \dots, \xi_{s})$. 
Then, we claim that smooth mappings of superdomains $\rm U^{p+q, s} \rightarrow V^{p'+q', s'}$ are in 
a bijective correspondence with parity preserving homomorphisms $ B_{p',q',s'}(V) \rightarrow B_{p,q,s}(U)$ 
(one can easily modify the proof  taking into account the  restriction  on 
the range  of first $p$ (respectively $p'$) coordinates).  
Now let us give a definition of supermanifold. 

\proclaim Definition. A $(p\oplus q|s)$-dimensional supermanifold $S$ 
is a superspace which is glued together from
$(p\oplus q|s)$-dimensional superdomains by means of smooth mappings. \par

In more detail, we assume  that there exists a covering $\cal U$ of 
the body $S_{\bf R}$ of superspace $S$  with the property that a restriction of 
$S$ to any  $U\in {\cal U}$ which we denote as $S^{U}$ is isomorphic to the superdomain 
$U_{\alpha}^{p\oplus q|s}$. 
 Let $U$ and $V$ be any two intersecting sets from the   covering $\cal U$ and let 
$\phi_{U}: S^{U} \rightarrow  U_{\alpha}^{p\oplus q|s}, \phi_{V}: 
S^{V} \rightarrow  V_{\alpha}^{p\oplus q|s}$ be the 
corresponding isomorphisms. Denote by  $\tilde \phi_{U} , \tilde \phi_{V}$ the isomorphisms 
$S^{U\cap V} \rightarrow  (U\cap V)_{\alpha}^{p\oplus q|s}$   induced by $\phi_{U}$ and $ \phi_{V}$ respectively. 
Then we require that the mappings $(\tilde \phi_{U})^{-1}\tilde \phi_{V}$ are smooth.


Using theorem 2 one can give a definition of a $(p\oplus q|s)$-dimensional supermanifold along the lines of 
a conventional Berezin-Leites approach to supermanifolds via ringed spaces. In those terms,  
a $(p\oplus q|s)$-dimensional supermanifold is a manifold ${\cal M}_{0}$  with a sheaf $\cal O$ of supercommutative 
rings on it with the property that locally over a neighborhood $U$ the sheaf is isomorphic to the algebra 
$B_{p,q,s}(U)= C^{\infty}({U})\otimes {\bf R}[[n_{1}, \dots , n_{q}]]\otimes \Lambda^{.}(\xi_{1}, \dots, \xi_{s})$. 


On the other hand, given a supermanifold $\cal M$ one can construct a sheaf of $Z_{2}$-graded supercommutative 
algebras over the body ${\cal M}_{0}$ as follows. For any open subset $U\subset {\cal M}_{0}$ consider 
the space of mappings  from the restriction of $\cal M$ to $U$ into ${\bf R}^{1|1}$. Denote this 
space by ${\cal F}(U)$. 
Note that one can identify ${\bf R}^{1|1}_{\Lambda}$ with $\Lambda$. This induces a structure of $Z_{2}$-graded 
supercommutative algebra on ${\cal F}(U)$ . It is easy to see that the collection of ${\cal F}(U)$ is a (pre)sheaf 
 endowing $\cal M$ with a structure of ringed space.     
The fact that locally our supermanifold is isomorphic to a superdomain determines the standard local 
structure of the  sheaf at hand.  

We define a tangent bundle $T{\cal M}$ of a supermanifold $M$ as a functor determined by 
$\Lambda$-points 
$(T{\cal M})_{\Lambda}=T({\cal M}_{\Lambda})$ and mappings $\tilde \alpha_{*}$ corresponding 
to morphisms $\alpha: \Lambda \rightarrow \Lambda'$. One can easily check that $T{\cal M}$ is 
a supermanifold itself and has  a canonical projection $T{\cal M} \rightarrow {\cal M}$.


It is possible to consider superspaces modelled on a fixed infinite-dimensional topological linear superspace $V$. 
  Such  a superspace $V$ can be constructed as in Example 2 out of any infinite-dimensional 
topological $Z_{2}$-graded linear space. An infinite-dimensional superdomain is a restriction of $V$ to 
any open subset of its body. In the infinite-dimensional case one can also define objects analogous to 
$(k\oplus l|q)$-dimensional linear superspaces. From now on 
let us fix $V$.  
One can give different definitions of differentiable (smooth) mappings between two topological 
linear spaces. For our purposes the following definition will be sufficient. 
We will call a mapping $F: V_{1} \rightarrow V_{2}$ 
between two infinite-dimensional topological linear superspaces differentiable  if  there exists a 
mapping $D:  V_{1}\times V_{1} \rightarrow V_{2}$ such that $D$ is linear in the second variable and 
for each AN algebra $\Lambda$ 
$F_{\Lambda}(x+h)=F_{\Lambda}(x) + D_{\Lambda}(x, h) + o_{x}(h)$ for any $x,h\in (V_{1})_{\Lambda}$. 
Here  $ o_{x}(h)$ vanishes to an order higher than $h$, i.e. for a real number $t \to 0$ , 
$t^{-1}o_{x}(th) \to 0$. 
(this is what is called differentiability in  the sense 
of $\rm G{\hat a}teaux$ , see for example \cite{ASbook}). 
Once we defined a differentiable mapping between  superdomains in $V$ we can consider 
infinite-dimensional differentiable supermanifolds modelled on $V$ as superspaces pasted from superdomains 
in $V$ by means of differentiable mappings.  


\smallskip 
{\it Lie supergroups}. We have already given above the definitions of supergroup and supermanifold. 
Following the same category-theoretic point of view we define a $(p\oplus q| s)$-dimensional 
Lie supergroup as a $(p\oplus q| s)$-dimensional supermanifold which is also a supergroup and for each 
AN algebra $\Lambda$ the set of $\Lambda$-points is a Lie group with respect to the given 
smooth manifold and group structures. To any $(p\oplus q| s)$-dimensional 
Lie supergroup one assigns naturally its Lie superalgebra being a  $(p\oplus q| s)$-dimensional vector 
space. All this can be  generalized easily to include  infinite-dimensional Lie supergroups modelled 
on a linear superspace $V$.

Given any Lie superalgebra $\cal G$ (possibly infinite-dimensional) one can consider its restriction to a 
point $0$ of the body and obtain a Lie superalgebra ${\cal G}_{nilp}$ which has only nilpotent coordinates. Then, 
 for each AN algebra $\Lambda$ one can consider  a set of formal power series 
$G_{\Lambda}= \{ exp(a) | a\in ({\cal G}_{nilp})_{\Lambda} \}$. 
The set $G_{\Lambda}$ is endowed with a Lie supergroup 
structure via the Campbell-Hausdorf formula which is represented by a finite sum because of nilpotency. 
It is easy to check that the sets $G_{\Lambda}$ constitute a    Lie supergroup $G_{nilp}$ modelled on a linear 
superspace $V={\cal G}_{nilp}$
and having ${\cal G}_{nilp}$ as its Lie superalgebra. 
If there exists a Lie supergroup $G$ having $\cal G$ as its Lie superalgebra, then $G_{nilp}$ can alternatively 
be constructed as the restriction of superspace $G$ to the subset of its body $ G^{(0)}$ consisting of one point 
-   the unit element of  Lie group $G^{(0)}$. These two constructions are canonically isomorphic 
( the isomorphism is set up via  exponential mappings in $G_{\Lambda}$'s).
In a more general situation  it  might happen that ${\cal G}$ contains an invariant Lie
subalgebra  ${\cal G}_{int} \subset {\cal G}$ such that there exists a Lie supergroup $G_{int}$ 
 with ${\cal G}_{int}= Lie (G_{int})$ (the subscript ``$int$" comes from the word ``integrable").
 Then, the construction above   can be  improved    as follows. 

 First note that the supergroup $G_{int}$ acts on $G_{nilp}$ by the adjoint action on ${\cal G}_{nilp}$. 
 Then we can consider a semidirect product of Lie supergroups $G_{nilp}\times_{Ad} G_{int}$ , i.e. the direct 
product supermanifold $G_{nilp}\times G_{int}$ with a group operation defined by the following 
formula $$ \left( e^{x},a \right) (e^{y}, b) = (e^{x} (a e^{y} a^{-1}), ab) $$ 
  As it was explained $G_{int}$ contains $(G_{int})_{nilp}$ which is canonically isomorphic to 
a Lie subgroup  $(G_{int})_{nilp} \subset G_{nilp}$. It is straightforward to check that the 
set of pairs $H= \{ (e^{x}, e^{-x}) \in G_{nilp}\times G_{int} | e^{x} \in (G_{int})_{nilp} \}$ is an 
invariant Lie subgroup of $G_{nilp}\times_{Ad} G_{int}$. 
 Taking the quotient with respect to $H$ 
we get a Lie supergroup whose Lie superalgebra is isomorphic to the one 
 obtained as a restriction of 
 Lie superalgebra $\cal G$ to a subset of its body ${\cal G}_{int}^{(0)} \subset {\cal G}^{(0)}$.


\smallskip
{\it Superspaces of maps.}
In this section we introduce  a superspace of maps and give some explicit constructions of this object. 
Let $\cal E, F$ be superspaces. 
For any superspace ${\cal S}$ a  mapping $\sigma:{\cal E}\times {\cal S} \rightarrow {\cal F}$ is called a 
family of mappings from the superspace $\cal E$ into the superspace $\cal F$ with base $\cal S$. 
A family of mappings $\nu: {\cal E}\times {\cal F^{E}} \rightarrow {\cal F}$ is called 
a universal family of mappings ${\cal E} \rightarrow {\cal F}$ if it has the following property . 
For any family ${\cal E}\times {\cal S} \rightarrow {\cal F}$ 
there exists a unique mapping $\rho : {\cal S} \rightarrow {\cal F^{E}}$ such that 
$\nu \circ (id_{\cal E}\times \rho) = \sigma$. In the case when $\cal E, F$ are supermanifolds we assume that all mappings 
are smooth and the base $\cal S$ of the family is a supermanifold though the base of the universal family 
need not be a supermanifold.   
For any particular pair of superspaces $\cal E, F$ there arises naturally
 a question of whether such a universal family of mappings exists. 
Provided a universal 
family exists and  is unique up to an isomorphism, 
 the base of  the universal family  is said to be the superspace of mappings   
(between  corresponding superspaces). Below we will construct  superspaces of maps
  between superdomains and 
describe a superspace of maps for the case of
 supermanifolds.

{\bf Example.} In this example we give a construction of the superspace of smooth maps 
${\bf R}^{p|q} \rightarrow {\bf R}^{p'| q'}$. Due to  theorem 1 from section 3 the smooth mappings 
${\bf R}^{p|q} \rightarrow {\bf R}^{p'| q'}$ are in a bijective correspondence with parity preserving 
homomorphisms $B_{p',q'} \rightarrow B_{p,q}$ ($B_{p,q}\equiv B_{p,q,0}$). These homomorphisms 
form a vector space $V^{p,q}_{p',q'}$ whose elements are rows of $p'$ even and $q'$ odd 
elements from $B_{p,q}$. Likewise one can consider a vector space $\tilde V^{p,q}_{p',q'}$ of 
parity reversing homomorphisms . Its elements are rows of $p'$ odd and $q'$ even elements from
$B_{p,q}$. Define a vector superspace ${\bf V}^{p,q}_{p',q'}$ as a vector superspace associated with 
a $Z_{2}$-graded vector space $V^{p,q}_{p',q'}\oplus \tilde V^{p,q}_{p',q'}$. We assert that   
${\bf V}^{p,q}_{p',q'}$ can be considered as a superspace of smooth mappings 
${\bf R}^{p|q} \rightarrow {\bf R}^{p'| q'}$.  To an element $b\in {\bf V}^{p,q}_{p',q'}(\Lambda)$ 
we assign a smooth mapping ${\bf R}^{p|q}_{\Lambda}\rightarrow {\bf R}^{p'|q'}_{\Lambda}$ in a natural way. 
One can check that this gives a family of smooth mappings 
$\nu^{p,q}_{p',q'}: {\bf R}^{p|q} \times {\bf V}^{p,q}_{p',q'} \rightarrow {\bf R}^{p'|q'}$.
Now let $\sigma: {{\bf R}^{p|q}}\times S \rightarrow {\bf R}^{p'| q'}$ be a family of 
smooth mappings. The most important case is    $S={\bf R}^{m|n}$. Then, 
${{\bf R}^{p|q}}\times S \cong {{\bf R}^{p+m| q+n}}$ and by Theorem 2 , $\sigma$ corresponds to an 
 element from $V^{p+m,q+n}_{p',q'}$. This element can be considered as a row of $p'+q'$  elements 
$\sigma^{k}, k=1,2,\dots, p'+q'$ from 
$B_{p+m,q+n}$ with the corresponding parity. Explicitly


$$
\sigma^{k}=\sum_{i=0}^{q} \sum_{j=0}^{n} \sum_{\alpha, \beta}
f^{k}_{\alpha_{1}, \cdots , \alpha_{i},\beta_{1},\dots,\beta_{j}}(x^{1},\dots,x^{p},y^{1},\dots, y^{m}) 
\xi^{\alpha_{1}}\dots \xi^{\alpha_{i}}\eta^{\beta_{1}}\dots\eta^{\beta_{j}} 
$$
where we assume that  variables $x^{i},\xi^{j}$ ( $y^{i}, \xi^{i}$)  are assigned to 
the superspace ${\bf R}^{p|q}$ (respectively ${\bf R}^{m|n}$) and the summation runs over the numbers 
$i$ and $j$ such that $i+j$ is of the appropriate parity . 
The collection of elements from $B_{p,q}$
 
\begin{eqnarray*}
g^{k}(x^{1},\dots,x^{p} )_{\alpha_{1}, \cdots , \alpha_{i}}&=&
\sum_{j=0}^{n} \sum_{1 \le \beta_{1} < \cdots < \beta_{i} \le n} 
f^{k}_{\alpha_{1}, \cdots , \alpha_{i},\beta_{1},\dots,\beta_{j}}(x^{1},\dots,x^{p},y^{1},\dots, y^{m})\\
&&\eta^{\beta_{1}}\dots\eta^{\beta_{j}} 
 \end{eqnarray*}
labeled by the sequence of multiindices $1\le \alpha_{1}<  \dots < \alpha_{i} \le q$, integer $k$, and the variables 
$x^{1},\dots, x^{p}$ defines a mapping 
$\rho: {{\bf R}^{m|n}} \rightarrow {\bf V}^{p,q}_{p',q'}$.  
One should simply utilize the standard substitutions of $\Lambda$-point  coordinates 
 into $g^{k}(x^{1},\dots,x^{p} )_{\alpha_{1}, \cdots , \alpha_{i}}$ and then form a combination 
$$\sum_{i=0}^{q}  \sum_{1 \le \alpha_{1} < \cdots < \alpha_{i} \le q} 
\xi^{\alpha_{1}}\dots \xi^{\alpha_{i}}g^{k}(x^{1},\dots,x^{p} )_{\alpha_{1}, \cdots , \alpha_{i}}
\in{\bf V}^{p,q}_{p',q'} (\Lambda)$$
A direct check shows that  
$\nu^{p,q}_{p',q'}\circ \rho=\sigma$. Using the fact that locally every supermanifold is equivalent to a 
superdomain   one can prove that ${\bf V}^{p,q}_{p',q'}$
 is indeed a superspace of 
maps. Note that the same  considerations as above go through for the case of mappings 
$U^{p|q}\rightarrow {\bf R}^{p'|q'}$ of a superdomain to a linear superspace. One simply needs to
 consider  vector spaces of parity preserving and parity reversing 
homomorphisms $B_{p',q'} \rightarrow B_{p,q}(U)$ and repeat the construction above. 
The superspace of maps between superdomains $U^{p|q}\rightarrow V^{p'|q'} $ is 
the restriction of superspace of maps $U^{p|q}\rightarrow {\bf R}^{p'|q'}$ to the subset of 
its body consisting of  maps carrying the body of the superdomain $U$ to the subset 
$V\subset {\bf R}^{p'|q'}$. The generalization of these results to the case of $(p\oplus q|s)$-dimensional 
superdomains is straightforward. One should replace everywhere algebras of the type $B_{p,q}$ 
by the algebras of the type $B_{p,q,s}$.

Now let us  discuss  briefly a situation in  the case of mappings between two supermanifolds. 
Let $\cal M$ and $\cal N$ be supermanifolds. 
Then, the restriction of superspace of maps ${\cal N}^{\cal M}$ 
to each connected component of its body can be endowed with a structure of infinite-dimensional 
supermanifold. For a given map $\phi :{\cal M} \rightarrow {\cal N}$ consider the pull-back of the 
tangent bundle $T{\cal N}$ induced by $\phi$. Denote this bundle as $\phi^{*} T{\cal N}$. 
The superspace of sections of $\phi^{*} T{\cal N}$ is a vector superspace  which can be regarded as a 
tangent space to ${\cal N}^{\cal M}$  at the point $\phi$ in a natural way. Putting some (super)Riemannian 
metric on $\cal N$ one can extend the infinitesimal variation of $\phi$ corresponding to a given 
section of  $\phi^{*} T{\cal N}$ to some finite variation. This construction determines a neighborhood 
   of the mapping $\phi$ being isomorphic to some open set (the restriction to some open subset of the body) 
of the vector superspace of sections $\Gamma(\phi^{*}T{\cal N})$.

 
\smallskip
{\it G-structures on supermanifolds.}
Our next goal is to introduce the notion of G-structure on  a supermanifold. 
Let $\cal M$ be a supermanifold of dimension $(m\oplus n|q)$ and let $\cal U$ be a covering of 
the body ${\cal M}_{R}$ such that the restriction of $\cal M$ to any $U\in {\cal U}$ is isomorphic to a 
superdomain $\rm U^{m\oplus n|q}$. For any $U,V\in {\cal U}$ denote by $\phi_{V,U}$ the corresponding 
gluing transformation (from the chart $U$ to $V$). We say that  
$L({\cal M})$ is  (the total space of) the frame bundle over $\cal M$ 
if $L({\cal M})$ is a supermanifold glued together from the superdomains 
$U^{m\oplus n|q}\times GL(m\oplus n|q)$ by means of 
transformations $\phi_{V,U}\times J$ where $J$ is the left multiplication by the jacobian matrix corresponding to 
$\phi_{U,V}$.  Explicitly, if $X=(x^{1}, \dots, x^{m}, y^{1}, \dots , y^{n}, \theta^{1}, \dots, \theta^{q})$ are 
coordinates on the superdomain $\rm U^{m\oplus n|q}$ then the gluing transformation from the 
 chart $U^{m\oplus n|q}\times GL(m\oplus n|q)$ to the chart  $V^{m\oplus n|q}\times GL(m\oplus n|q)$ acts on the element 
$(a, g)\in \rm U^{m\oplus n|q}_{\Lambda}\times GL_{\Lambda}(m\oplus n|q)$ as follows: 
$(a, g)\mapsto (\phi_{V,U}(a), \frac{\partial \phi_{V,U}}{\partial X} (a) g)$. The supermanifold $L({\cal M})$ 
constructed this way is equipped with a natural structure of principal $GL(m\oplus n|q)$bundle over $\cal M$. 
One can show that $L({\cal M})$ does not depend on the choice of covering $\cal U$.
Let $G$ be a Lie (super)subgroup of $GL(m\oplus n|q)$ then,  any smooth mapping ${\cal M} \rightarrow G$ acts  
 naturally from the right on  $L({\cal M})$. Furthermore, the superspace of maps $G^{\cal M}$ can be 
furnished with a natural supergroup structure (in a ``pointwise" manner). This supergroup acts on 
$L({\cal M})$. Taking the quotient space with respect to this action one obtains a superspace denoted by
$L({\cal M})/G^{\cal M}$. Since the elements of   $G^{\cal M}$ act fiberwise , the superspace  
$L({\cal M})/G^{\cal M}$ is 
equipped with a structure of fiber bundle over $\cal M$. 
For any Lie (super)subgroup $G\subset GL(m\oplus n|q)$ we say that there is a $G$-structure  defined on 
a supermanifold $\cal M$ if a section of the bundle $L({\cal M})/G^{\cal M} \rightarrow {\cal M}$ is specified. 
In general, given a bundle $p: {\cal E} \rightarrow {\cal B}$ where ${\cal E}, {\cal B}$ one  can consider the 
corresponding superspace of sections denoted by $\Gamma ( {\cal E} )$  as a subspace in $\cal E^{B}$. 
Thus we can consider the superspace of sections of the bundle $L({\cal M})/G^{\cal M} \rightarrow {\cal M}$. 
We call this superspace  the superspace of $G$-structures over $\cal M$. 
The supergroup of diffeomorphisms of $\cal M$ acts naturally on the superspace of $G$-structures over 
$\cal M$. 
The corresponding
quotient space  is called 
the moduli space of $G$-structures.  Note that this moduli space  is a superspace by definition but in 
general it is not a supermanifold. 
In particular situations the most interesting case is that of locally
standard $G$-structures.
For a $G$-structure to be locally standard we mean the following . 
Fix a $G$-structure $B_{st}$ on ${\bf R}^{m+n|q}$ that will be called standard. Note that  a $G$-structure 
on supermanifold $\cal M$ induces a $G$-structure on any superspace obtained by
 restriction to a subset of the body $U\subset {\cal M}_{R}$. Assume $\cal U$ is a standard atlas for the manifold 
$\cal M$, i.e. $\cal M$ restricted  to any element from $\cal U$ is equivalent to a superdomain. Then, a $G$ structure 
 $B$ on $\cal M$ is called locally standard if  the restriction of $B$ to any neighborhood $U\in {\cal U}$ 
is isomorphic to the restriction of   $B_{st}$ to $U$.

A rigorous definition of $(p,q)$-superconformal manifold can also be given in the language of 
$G$-structures. This would provide a rigorous definition for the moduli space of $(k\oplus l|q)$-dimensional 
supermanifolds  as was  outlined above for the general case of arbitrary $G$-structure.


\end{document}